\documentclass{article}
\usepackage[english]{babel}
\usepackage{amsfonts,amssymb, amsmath}

\newtheorem{prop}{Proposition}

\begin{document}

\title{Homogeneous potentials, Lagrange's identity and Poisson geometry}
\author{A.\,V.~Tsiganov\\
\it\small International Laboratory for Mirror Symmetry\\ 
\it\small National Research University Higher School of Economics, 
Moscow, Russia\\
\it\small Beijing Institute of Mathematical Sciences and Applications, Beijing, China\\
\it\small email: andrey.tsiganov@gmail.com} 

\date{}
\maketitle

\begin{abstract}
The Lagrange identity expresses the second derivative of the moment of inertia of a system of material points
through kinetic energy and homogeneous po\-ten\-ti\-al energy, from which follows the Jacobi well-known result on the instability of a system of gravitating bodies. In this work, it is proven that if a Hamiltonian system satisfies the Lagrange identity, then it possesses additional tensor invariants that are not expressed through the basic invariants existing for all Hamiltonian systems. A new class of Hamiltonian systems with in\-ho\-mo\-ge\-ne\-ous potentials is considered, which also possess similar additional tensor invariants. 
\end{abstract}

\section{Introduction}
\setcounter{equation}{0}
Let us consider the Hamiltonian system of differential equations
\begin{equation}\label{h-eq}
\frac{dq_i}{dt}= \frac{\partial H}{\partial p_i}\,,\qquad
\frac{dp_i}{dt}= -\frac{\partial H}{\partial q_i}\,,\qquad i=1,\ldots,n,
\end{equation}
where $q=\{q_1,\ldots,q_n\}$ are coordinates in configuration space $\mathbb R^n$,  $p=\{p_1,\ldots,p_n\}$ are the corresponding moments in phase space $T^*\mathbb R^n$, and $H=H(q,p)$ is the Hamiltonfunction. 

If the Hamilton function 
\begin{equation}\label{ham-1}
H=T(p)+V(q)=\frac{1}{2m_1} p_1^2+\cdots+\frac{1}{2m_n} p_n^2+V(q)\,,\qquad m_i>0\,,
\end{equation} 
is the sum of kinetic energy $T(p)$ and a smooth homogeneous function $V(q)$ of degree $k$,
then Lagrange's identity holds
\begin{equation}\label{eq-lag}
\frac{d^2 J}{dt^2}= 4T - 2kV=4H-2(k+2)V,\qquad J=\sum_{i=1}^n m_iq_i^2\,.
\end{equation}
Jacobi used  this identity considering the stability of the motion of systems
of $n$ interacting particles with masses $m_1,\ldots,m_n$ arising in celestial mechanics \cite{jac}, see also \cite{alb}. Various generalizations and applications of Lagrange's identity are discussed in the works of V.V. Kozlov \cite{koz,koz24,koz25}.  

The necessary conditions for the integrability of equations (\ref{h-eq}) in the class of meromorphic functions are discussed in the works \cite{adam,llibre}. Below we consider Hamiltonian systems of general position, i.e., potentials that are not subject to any conditions other than the condition of homogeneity
\[
V(\lambda\,q_1,\ldots,\lambda\,q_n) =\lambda^k\,V(q_1,\ldots,q_n)\,.
\]
or, according to the Euler theorem for smooth homogeneous functions, the condition 
\begin{equation}\label{euler}
Z=\sum_{i=1}^n q_i\frac{\partial}{\partial q_i}\,V(q_1,\ldots,q_n)-kV(q_1,\ldots,q_n)=0\,,
\end{equation}
which is equivalent to Lagrange's identity (\ref{eq-lag}) in the context of Hamiltonian mechanics.

Equations (\ref{h-eq}) can be interpreted geometrically in terms of the vector field $X$ on the phase space $T^*\mathbb R^n$, which in local coordinates $x=(q,p)$ has the form
\begin{equation}\label{x-vec}
X=\sum_{i=1}^n \left( \frac{p_i}{m_i} \frac{\partial}{\partial q_i}- \frac{\partial V}{\partial q_i}\dfrac{\partial}{\partial p_i}\right)\,.
\end{equation}
Using the vector field $X$, Hamilton's original equations (\ref{h-eq}) are rewritten in an invariant geometric form
\begin{equation}\label{h2-eq}
\dot{x}=X\,,\quad\mbox{where}\quad X=P dH\qquad \mbox{or}\qquad \iota_X \omega=-dH\,,
\end{equation}
 which contains the following basic tensor invariants: the Hamilton function $H$, the Poisson bivector $P$, and the symplectic form $\omega$, which in local coordinates have the form 
\begin{equation}\label{p-can}
P=\sum_{i=1}^n\frac{\partial}{\partial q_i}\wedge\frac{\partial}{\partial p_i}\qquad\mbox{and}\qquad
\omega=\sum_{i=1}^n dp_i\wedge dq_i\,.
\end{equation}
The corresponding Poisson bracket on the phase space $T^*\mathbb R^n$ has a canonical form
\[
\{q_i,p_j\}=\delta_{ij}\,,\qquad \{q_i,q_j\}=\{p_i,p_j\}=0\,,\qquad i,j=1,\ldots,n.
\]

In this note we prove that if potential $V(q)$ is a smooth homogeneous function of degree $k$, then for the Hamiltonian system (\ref{h-eq},\ref{h2-eq}) not only Lagrange's identity (\ref{eq-lag}) holds, but also exist an invariant Poisson bivector $\hat{P}$ and an invariant sym\-plec\-tic form $\hat{\omega}=\hat{P}^{-1}$, which cannot be obtained by differentiation and tensor operations from the basic tensor invariants $X,H,P$ and $\omega$.

\subsection{Basic tensor invariants of Hamiltonian systems}
Recall that  smooth tensor field $T(x)$ is called a tensor invariant of system of ordinary differential equations (\ref{h2-eq}) if 
\begin{equation}\label{g-inv}
\mathcal L_XT(x) = 0\,.
\end{equation}
Here, $\mathcal L_X$ is the Lie derivative of the tensor field $T(x)$ along the vector field $X$, which determines the rate of change of the tensor field $T(x)$ during deformation of the phase space, defined by the phase flow of the system (\ref{h2-eq}). 
 
 The vector field $X$ is a tensor invariant due to the antisymmetry of the Lie brackets $[.,.]$ 
\[
\mathcal L_X\,X=[X,X]=0\,.
\]
Substituting the Hamiltonian vector field $X=PdH$ into this equation, we obtain
$X=PdH$
\begin{equation}\label{sum-inv}
\mathcal L_X\,X=\mathcal L_X\,\left(PdH\right)=\left(\mathcal L_X P\right)dH+Pd\left(\mathcal L_X H\right)=0\,.
\end{equation}
Both terms in this identity are equal to zero due to the definition of the Poisson bivector $P$, which is a skew-symmetric tensor field of valence $(2,0)$ (bivector)
\[
\mathcal L_X\,H=(dH,PdH)=0\,,
\]
which is invariant of the phase flow  
\begin{equation}\label{inv-gen}
\mathcal L_X\,P=\mathcal L_{PdH}\,P=[\![P,P]\!] dH=0
\end{equation}
by virtue of the Jacobi identity 
\begin{equation}\label{jac-eq} 
[\![P,P]\!]=0\,,
\end{equation}
 which we write using the Schouten brackets $[\![.,.]\!]$. 

It is obvious that Jacobi's identity (\ref{jac-eq}) is a sufficient but not necessary condition for satisfying the invariance equation (\ref{inv-gen}). For example, the bivector 
\[P'=(a_1H+a_2)P\,,\qquad a_{1,2}\in\mathbb R\,,\] 
does not satisfy Jacobi's identity, but nevertheless is a tensor invariant for arbitrary values of the parameters $a_{1,2}$. 

The invariance of the symplectic form is a consequence of condition $d\omega=0$ and Hamilton's equations (\ref{h2-eq}), since
\[
\mathcal L_X \omega=d(\iota_{ X}\omega) + \iota_{X}(d\omega) = d(-dH) = 0\,,
\]
which leads to the preservation of the $2n$ volume form
\[
\Omega=\wedge^n \omega=(-1)^{n(n+1)/2}\,n!\, dq_1\wedge\cdots\wedge dq_n\wedge dp_1\cdots\wedge dp_n\,,\qquad \mathcal L_X\Omega=0\,.
\]
According to Cartan's  formula, the Lie derivative $\mathcal L_X$ acting on the differential $k$-form $\varphi$ is the anticommutator of the inner product with the outer derivative
\begin{equation}\label{hom-eq}
\mathcal L_X\varphi=\iota_X d\varphi+d\iota_X\varphi\,.
\end{equation}
Here, $\iota_X\varphi$ is the inner product of the vector $X$ and the form $\varphi$, that is, the $(k - 1)$-form obtained from the form $\varphi$ after substituting the vector field $X$ for the first argument. 

According to Poincar\'{e} \cite{poi} and Cartan \cite{car}, the differential form $\varphi$ is an absolute integral invariant if both terms in (\ref{hom-eq}) are simultaneously equal to zero.
\[
\iota_X d\varphi=0\qquad\mbox{and}\qquad d\iota_X\varphi=0\,.
\]
These conditions are sufficient but not necessary for satisfying the invariance condition  $\mathcal L_X\varphi=0$ with the Lie derivative given by (\ref{hom-eq}) and, therefore, in works \cite{poi,car}, conditional integral invariants were introduced in addition to absolute invariants, see also discussion in \cite{koz19,koz25a}.

Absolute and conditional tensor invariants for some integrable and non-integrable systems have been studied in \cite{ts25,ts25a,ts25b,ts25c}. In the next section, we consider a generalization of the results obtained in \cite{ts25c} to the case of arbitrary dimension of the phase space $T^*\mathbb R^n$. 

\subsection{Homogeneous vector fields}
Let us consider smooth vector field 
\[
X=\sum_{i=1}^n X_i(x_1,\ldots,x_n)\frac{\partial}{\partial x_i} 
\]
whose elements $X_i(x_1,\ldots,x_n)$ are homogeneous functions of the variables $x_1,\ldots,x_n$ with homogeneity index $\alpha$
\[
X_i(\lambda x_1,\ldots,\lambda x_n)=\lambda^\alpha X_i(x_1,\ldots, x_n)\,,\qquad i=1,\ldots,n\,.
\]
According to Euler's theorem on differentiable homogeneous functions, the Lie bracket between the vector field $X$ and the Euler vector field $E$ is equal to
\[
[E,X]=(1-\alpha)X\,,\qquad 
E=\sum_{i=1}^n x_i\frac{\partial}{\partial x_i}\,.
\]
It follows, therefore, that the bivector
\begin{equation}\label{my-p}
P_e =X\wedge E
\end{equation} 
is a tensor invariant of the phase flow defined by the vector field $X$
\[\mathcal L_X\,P_e =[X,X]\wedge E+X\wedge [X,E]=0\,.\]
which satisfies to the Jacobi condition 
\[
[\![P_e,P_e]\!]=2X\wedge [E,X]\wedge E=0\,.
\] 
Suppose that the function $F(x_1,\ldots,x_n)$ is a homogeneous first integral of the vector field $X$ with homogeneity index $\kappa$, i.e.
\[
X(F)=\sum_{i=1}^n X_i\frac{\partial F}{\partial x_i}=0\qquad\mbox{and}\qquad
E(F)=\sum_{i=1}^n x_i\frac{\partial F}{\partial x_i}=\kappa F\,.
\]
In this case we have
\[
P_edF=-X(F)\,E+E(F)\,X=\kappa F\, X\,,
\]
that allows us to represent a homogeneous vector field $X$ in formal Hamiltonian form
\[
X=P_edH\,,\qquad H=\kappa^{-1}\ln |F|\,.
\]
Here, $P_e$ is the second-rank Poisson bivector (\ref{my-p}), and $H$ is the Hamiltonian. 

According to \cite{bizkoz}, we  call the vector field $X=P_edH$ a formal Hamiltonian field because the Poisson bivector $P_e$ (\ref{my-p}) does not possess $m=n-2$ uniquely determined and independent Casimir functions $C_1,\ ..., C_m$ such that $P_edC_i=0$, $i=1,\ldots,m$. 

\section{Homogeneous potentials}
\setcounter{equation}{0}
In the general case, the vector field $X$ (\ref{x-vec}) arising in Hamiltonian mechanics is not homogeneous. However, if the potential $V(q)$ is a smooth homogeneous function, then the Hamiltonian $H=T+V$ (\ref{ham-1}) consists of two homogeneous functions $T$ and $V$, which allows us to construct an analogue of the invariant Poisson bivector (\ref{my-p}). 

Indeed, let us define an Euler-type vector field  
\begin{equation}\label{y-1}
Y=\sum_{i=1}^n q_i\frac{\partial}{\partial q_i}+\frac{k}{2}
\sum_{i=1}^n p_i\frac{\partial}{\partial p_i}\,,
\end{equation}
for which the sum of two homogeneous functions $H=T+V$ (\ref{ham-1}) will be a Darboux invariant \cite{d78a}
\[
\mathcal L_Y H\equiv Y(H)=k H
\]
provided that the condition $Z=0$ (\ref{euler}) holds. The corresponding vector field $X=PdH$ (\ref{x-vec}) will also be a Darboux invariant of the vector field $Y$
\begin{equation}\label{y-eq}
\mathcal L_YX \equiv [Y,X]=\left(\frac{k}{2}-1\right)X+PdZ\,,\qquad\mbox{where}\qquad Z=0\,.
\end{equation}
Analogous to the bivector $P_e=X\wedge E$ (\ref{my-p}) considered in the previous section, the bivector 
\begin{equation}\label{p-p}
P'=X\wedge Y\,,
\end{equation}
is invariant with respect to the flow of the vector field $X$
\[
\mathcal L_X P'=[X,X]\wedge Y+X\wedge [X,Y]=0
\]
and satisfies the Jacobi condition
\[
[\![P',P']\!]=2X\wedge [Y,X]\wedge Y=0\,.
\] 
which is a direct consequence of equation (\ref{y-eq}), valid only when the Lagrange identity (\ref{eq-lag}) holds.

We have thus proven the following statement.
\begin{prop}
If the Hamiltonian function $H=T(p)+V(q)$ (\ref{ham-1}) is a Darboux invariant of the vector field $Y$ (\ref{y-1})
\[Y(H)=k H\qquad\mbox{at}\qquad H\neq 0\,,\] then the tensor field $P'$ (\ref{p-p}) is a Poisson bivector invariant with respect to the phase flow generated by the Hamiltonian vector field $X$ (\ref{x-vec}).
\end{prop}
In terms of the variables $q$ and $p$, this Poisson bivector takes the following form 
\begin{align*}
P'=&\sum_{1\leq i<j\leq n}\left(\frac{p_iq_j}{m_i}-\frac{p_jq_i}{m_j}\right)\frac{\partial}{\partial q_i}\wedge\frac{\partial}{\partial q_j}+ \sum_{i=1}^n\sum_{j=1}^n
\left(\frac{k}{2m_i}p_ip_j +q_i\frac{\partial V}{\partial q_j}\right)\frac{\partial}{\partial q_i}\wedge\frac{\partial}{\partial p_j}
\nonumber\\
+&\sum_{1\leq i<j\leq n} \frac{k}{2}
\left(
p_i\frac{\partial V}{\partial q_j}-p_j\frac{\partial V}{\partial q_i}
\right)\frac{\partial}{\partial p_i}\wedge\frac{\partial}{\partial p_j}\,.
\end{align*}
For $k \neq -2$, the invariant Poisson bivector $P'$ (\ref{p-p}) does not compatible with the Poisson bivector $P$ (\ref{p-can}) since
\[[\![P,P']\!] \neq 0.\]
Recall that the Poisson bivectors $P$ and $P'$ are compatible if any linear combination $(a P+b P')$ with constant coefficients $a$ and $b$ is a Poisson tensor, that is, the Jacobian identity holds
\[
[\![a P+b P',a P+b P']\!]=2ab[\![P,P']\!]=0\,,\qquad \forall a,b\in \mathbb C.
\]
In the case of the Jacobi potential field for $k=-2$, the compatibility condition holds: $[\![P,{P}']\!]=0$, and in this case
there is another scalar invariant
\[
F=\frac{1}{2}\sum_{i,j=1}^n (q_ip_j-q_jp_i)^2 +2 V\sum_{i=1}^nq_i^2\,,\qquad \mathcal L_XF=0\,,
\]  
see the definition of the Jacobi potential field and the corresponding discussion in \cite{koz25}.

The Poisson bivector $P'$ (\ref{p-p}) is degenerate for $n\geq2$, since $P'$ has rank 2, and satisfies the relation
\[
P'dH=kHX\,.
\]
In the general case, the Poisson bivector $P'$ (\ref{p-p}) does not compatible with the Poisson bivector ${P}$ (\ref{p-can}). Nevertheless, there exists a unique linear combination of these invariant bivectors with coefficients depending on the Hamiltonian function $H$ and the degree of homogeneity of the potential $k$, which is an invariant Poisson tensor.  
\begin{prop}
If $V(q)$ is a homogeneous function of degree $k \neq \pm2$, then on the open subset defined by the inequality $H \neq 0$,
the tensor field of valence $(2,0)$
\begin{equation}\label{p-hat}
\hat{P}=\left(1 - \frac{k}{2}\right)H P + P'\,,\qquad \mbox{rank}\hat{P}=n\,,
\end{equation}
is a non-degenerate Poisson bivector. This allows us to define on  this open subset a tensor field of valence $(0,2)$
\[
\hat{\omega}=\hat{P}^{-1}
\]
which is a symplectic form invariant with respect to the phase flow generated by the Hamiltonian vector field $X$ (\ref{x-vec}).
\end{prop}
The proof consists of directly verifying the non-degeneracy, invariance, and the Jacobi identity
\[\mathcal L_X\hat{P}=0\,,\qquad [\![\hat{P},\hat P]\!]=0\qquad\mbox{and}\qquad \mathcal L_X\hat{\omega}=0\,,\qquad d\hat{\omega}=0\]
which are equivalent to the condition   $Z=0$ (\ref{euler}).

According to Lie-Darboux theorem, there exists a coordinate transformation $\phi:(q,p)\to(\hat{q},\hat{p})$ such that
\[
\hat{P}=\sum_{i=1}^n\frac{\partial}{\partial \hat{q}_i}\wedge\frac{\partial}{\partial \hat{p}_i}\qquad \mbox{and}\qquad
\hat{\omega}=\sum_{i=1}^n d\hat{p}_i\wedge d\hat{q}_i
\]
at least locally. This transformation can be regarded as an analogue of the Baecklund transformations for integrable Hamiltonian systems with homogeneous potentials \cite{ts15}.

For all three invariant Poisson bivectors $P$, $P'$, and $\hat{P}$, the Hamiltonian flows generated 
by the Hamiltonian function $H$ with potential $V(q)$ (\ref{euler}) are proportional to each other 
\[P'dH=kHX\qquad \mbox{and} \qquad  \hat{P}dH=\frac{k+2}{2}H\,X\]
where $X=PdH$.

We note that the cases $k=\pm 2$ are special cases within the application of Galois group theory to Hamiltonian systems with homogeneous potentials \cite{gal}.

\section{Nonhomogeneous potentials of a special kind}
\setcounter{equation}{0}
Let us consider the Hamilton function
\[
H=\frac{1}{2m_1} p_1^2+\cdots+\frac{1}{2m_n} p_n^2+U(q)
\]
where the potential $U(q)$ satisfies the relation 
\begin{equation}\label {euler2}
\bar{Z}=0\,,\qquad\mbox{where}\qquad \bar{Z}=\sum_{i=1}^n a_i\frac{\partial}{\partial q_i}\,U(q_1,\ldots,q_n)-kU(q_1,\ldots,q_n)\,.
\end{equation}
In this case, the equivalent of Lagrange’s identity (\ref{eq-lag}) is the expression
\begin{equation}\label{eq-lag2}
\frac{d^2}{dt^2} \sum_{i=1}^n a_iq_i = kU.
\end{equation}
The well-known particular solution of Euler's equation (\ref{euler}) has the following  form  
\[
V(q_1,\ldots, q_n)=q_1^kf\left(\frac{q_2}{q_1},\ldots,\frac{q_n}{q_1}\right)
\]
where $f$ is an arbitrary smooth function of projective coordinates. Similar particular solution of equation (\ref{euler2}), has the form
\[
U(q_1,\ldots, q_n)=e^{\frac{k}{a_1}q_1}g\left(
q_2-\frac{a_2}{a_1}\,q_1,\ldots, q_n-\frac{a_n}{a_1}\,q_1\right)\,,
\]
where $g$ is an arbitrary smooth function. A special case of Hamiltonian systems with potential $U(q)$ are some of the Toda lattices, see \cite{ts25c}. 

For the vector field 
\begin{equation}\label{y-2}
\bar{Y}=\sum_{i=1}^n a_i\frac{\partial}{\partial q_i}+\frac{k}{2}
\sum_{i=1}^n p_i\frac{\partial}{\partial p_i}\,,
\end{equation}
our Hamiltonian $H=T+U$ is the Darboux invariant at $\bar{Z}=0$ since
\[
\bar{Y}(T+U)=k(T+U)+Pd\bar{Z}\,.
\]
The corresponding vector field
 \[
\bar{X}=PdH=\left(\sum_{i=1}^n \frac{p_i}{m_i} \frac{\partial}{\partial q_i}\right)-
\left(\sum_{j=1}^n \frac{\partial U}{\partial q_j}\dfrac{\partial}{\partial p_j}\right)
\]
is also a Darboux invariant for the vector field $\bar{Y}$ (\ref{y-2})
\[
[\bar{Y},\bar{X}]=\left(\frac{k}{2}-1\right)\bar{X}\qquad\mbox{at}\qquad \bar{Z}=0\,.
\]
From this, the following statement immediately follows.

\begin{prop}
Second-rank bivector
\[\bar{P}'=\bar{X}\wedge\bar{Y}\]
is  the invariant Poisson bivector for the phase flow of the vector field $\bar{X}$ if the function $U(q)$ satisfies equation (\ref{euler2}), that is, if
\[\bar{Y}(H)=kH\qquad\mbox{at}\qquad H\neq 0\,.\]
\end{prop}
The proof consists of substituting the Lie bracket between the vector fields 
$\bar{X}$ and $\bar{Y}$ into the equations 
\[\mathcal L_{\bar{X}}\,\bar{P}' =[\bar{X},\bar{X}]\wedge \bar{Y}+\bar{X}\wedge [\bar{X},\bar{Y}]=0\,.\]
and 
\[
[\![\bar{P}',\bar{P}']\!]=2\bar{X}\wedge [\bar{Y},\bar{X}]\wedge \bar{Y}=0\,.
\]
In terms of the variables $q$ and $p$, this Poisson bivector takes the following form 
\begin{align}
\bar{P}'=&\sum_{1\leq i<j\leq n}\left(\frac{a_jp_i}{m_i}-\frac{a_ip_j}{m_j}\right)\frac{\partial}{\partial q_i}\wedge\frac{\partial}{\partial q_j}+
\sum_{i=1}^n\sum_{j=1}^n
\left(\frac{k}{2m_i}p_ip_j +a_i\frac{\partial U}{\partial q_j}\right)\frac{\partial}{\partial q_i}\wedge\frac{\partial}{\partial p_j}\nonumber\\
+&\sum_{1\leq i<j\leq n} \frac{k}{2}
\left(
p_i\frac{\partial U}{\partial q_j}-p_j\frac{\partial U}{\partial q_i}
\right)\frac{\partial}{\partial p_i}\wedge\frac{\partial}{\partial p_j}\,.
\label{p-p2}
\end{align}
The Poisson bivector $\bar{P}'$ (\ref{p-p2}) is degenerate for $n\geq2$, since $\bar{P}'$ has rank 2, and satisfies the relation
\[
P'dH=kHX\,.
\]
In the general case, the Poisson bivector $\bar{P}'$ (\ref{p-p2}) is not commutative with the Poisson bivector ${P}$ (\ref{p-can}). Nevertheless, as in the case of homogeneous potentials, there exists a unique linear combination of these invariant bivectors with coefficients depending on the Hamiltonian $H$ and the parameter $k$, which is the invariant Poisson tensor.
\begin{prop}
If the potential $U(q)$ satisfies the relation (\ref{euler2}) for $k \neq 0$, then on the open subset defined by the inequality $H \neq 0$, the tensor field of valence $(2,0)$
\begin{equation}\label{p-hat2}
\bar{P}= - \frac{k}{2} H P + \bar{P}'\,,\qquad \mbox{rank}\bar{P}=n\,,
\end{equation}
is a non-degenerate invariant Poisson bivector, and the $(0,2)$-valence tensor field
\[
\bar{\omega}=\bar{P}^{-1}
\]
is a symplectic form invariant under the phase flow generated by the Hamiltonian vector field $\bar{X}$.
\end{prop}
The proof consists of a direct verification.

According to the Lie-Darboux theorem, there exists a canonical transformation of variables  $\phi:(q,p)\to(\bar{q},\bar{p})$ such that
\[
\bar{P}=\sum_{i=1}^n\frac{\partial}{\partial \bar{q}_i}\wedge\frac{\partial}{\partial \bar{p}_i}\qquad \mbox{and}\qquad
\bar{\omega}=\sum_{i=1}^n d\bar{p}_i\wedge d\bar{q_i}\,.
\]
This transformation can be viewed a counterpart of the B{\"a}cklund transformations for integrable Hamiltonian systems.

For all three Poisson invariant vectors $P$, $\bar{P}'$, and $\bar{P}$, the Hamiltonian fluxes generated 
by the Hamiltonian function $H$ with potential $U(q)$ (\ref{euler2}) are proportional to 
\[\bar{P}'dH=kH\bar{X}\qquad \mbox{and} \qquad  \bar{P}dH=\frac{k}{2}H\bar{X}\]
where $\bar{X}=PdH$, i.e., they can be reduced to each other by a time substitution.

\section{Conclusion}
According to \cite{bog96,bog97}, integrable Hamiltonian systems possess a family of tensor invariants of valence $(2, 0)$ and $(0,2)$, which cannot be obtained from the basic tensor invariants using various tensor operations; see examples in 
\cite{ts25,ts25a,ts25b,ts25c}.

This paper proves the existence of additional tensor invariants for non-integrable Hamiltonian systems, provided that the Hamiltonian function $H$ is a Darboux invariant with respect to the Euler type vector fields, which is equivalent to the condition that the Lagrangian identity or its analogue holds. 

The existence of such tensor invariants for non-integrable systems raises a number of questions  about the properties of these invariant geometric structures, about the qualitative behavior of trajectories that preserve these geometric invariants, about what distinguishes these non-integrable systems from the rest, about the use of these invariants for numerical integration, and so on. We do not yet know the answers to these questions.

The author thanks A.J. Maciejewski and an anonymous reviewer for their extensive comments on the first version of this work, which allowed for its significant improvement

The article was prepared within the framework of the project "International academic cooperation" HSE University.

\end{document}